# A Scalable Methodology for Reinstating the Superhydrophilicity of Ambient-Contaminant Compromised Surfaces

Ilias Papailias, Arani Mukhopadhyay, Anish Pal, Shahriar Namvar, Constantine M. Megaridis[*]

Department of Mechanical and Industrial Engineering, University of Illinois Chicago, Chicago, IL 60607, USA

* Corresponding author e-mail: cmm@uic.edu

**ABSTRACT**

The degradation of the hemi-wicking property of superhydrophilic high-energy surfaces due to contaminant adsorption from the ambient atmosphere is well documented. This degradation compromises the performance of such surfaces, thus affecting their efficacy in real-world applications where hemi-wicking is critical. In this work, the role of surface micro/nanostructure morphology of laser-textured metallic surfaces on superhydrophilicity degradation is studied. We explore intrinsic contact angle variations of superhydrophilic surfaces via adsorption of organics from the surroundings, which brings about the associated changes in surface chemistry. Furthermore, we explore condensation from humid air as a scalable and environment friendly methodology that can reinstate surface superhydrophilicity to a considerable extent (64% recovery of intrinsic wettability after three hours of condensation) due to the efficient removal of physisorbed contaminants from the surface texture features. The present results strengthen the argument that contact-line movement at fine scales can be used for de-pinning and removal of adsorbed organic molecules from contaminated surfaces.

## 1. INTRODUCTION

Hemiwicking (i.e. capillary spreading of a liquid on a rough hydrophilic surface) -frequently observed in nature as well as in technological applications- is a combined result of surface micro/nanostructures and high surface energy of the underlying solid.[1] The wettability of a solid can be quantified by the contact angle of millimeter-diameter water droplets,[2] or in extreme cases (when the contact angle is too low) by measuring the liquid lateral spreading rate.[3]

Certain technological applications necessitate the continuous presence of a thin liquid film, as for example in enhanced heat transfer,[4] photovoltaic applications,[5] maintaining proper thermophysical boundary conditions,[6,7] enhanced drainage and collection of a dispersed phase,[8] tunable wetting topology,[9,10] formation of 2D amorphous clathrates,[11] etc. Maintaining a thin liquid film on boundary walls of chambers that explore aerosol-cloud-drizzle interactions in a turbulent environment can also be compromised by the degradation examined herein.[12,13] Therefore, in such applications the underlying role of the surface micro/nanotexture of high surface-energy materials critically influences not only the wicking rate of liquids (or hemiwicking)[14] but also the surface-chemistry interactions.[15,16] Among a plethora of methods capable of creating scalable textured surfaces that are also superhydrophilic,[17-22] laser ablation has recently gained popularity for efficient surface functionalization,[23-26] especially owing to its high microstructure precision, repeatability, and overall scalability.

During laser ablation, a large amount of energy is released at a prescribed frequency by a pulsed or continuous laser beam, thus creating patterns on (or selectively ablate) a surface. The texture and morphology of the surface patterns is controlled through the laser parameters (beam power, frequency, traversing speed, raster line spacing) that are set in advance of the laser-processing step. By adjusting these laser parameters, different surface architectures can be achieved with high spatial precision and control. As a result of its advantages, laser processing has found use in several technological applications, ranging from phase-change heat transfer and biology, to microfluidics.[27] However, the exceptional hydrophilicity and hemiwicking properties of these high-energy surfaces may degrade over time due to adsorption and accumulation of volatile organic compounds (VOCs) during ambient atmosphere exposure (down time); such adsorption eventually lowers the energy of the surface, thereby turning it hydrophobic, and in turn, compromising its hemiwicking performance.[28-31] The most common contamination sources are hydrocarbons that are ubiquitously present in ambient air.[32,33] These contaminants can be partially removed from the surface by washing with solvents (e.g. ethanol, acetone, isopropanol, deionized water), or even more effectively, by applying high temperatures (over 180 °C) and complex solvents.[34] However, these recovery methods are practically infeasible for large-scale applications, such as convection-cloud chambers, solar thermal desalinators, or falling film evaporators,[35] whose surfaces may extend to many square meters. Thus, new approaches that could maintain the continuous wicking functionality of superhydrophilic surfaces and the longevity of their application must be sought. At the same time, the recovery of superhydrophilicity of exposed surfaces has been rarely studied, as most works up to now have focused on the performance of hydrophobic and superhydrophobic surfaces.[36-40]

In the present work, the role of surface micro/nanostructure morphology on superhydrophilicity degradation is studied. First, stainless steel surfaces were enhanced with hierarchical microstructures through laser ablation, thus rendering them superhydrophilic. The surfaces were subsequently exposed to a research laboratory open environment, where their superhydrophilicity degraded due to the deposition of VOCs from the air. It has been established that nanofeatures formed on the microstructures formed during the laser processing step can prevent the accumulation of contaminants on a surface.[41] Next, we explored condensation from humid air as a new simple methodology that can reinstate superhydrophilicity and hemiwicking by removing the physisorbed contaminants. In contrast with other methods relying on the use of chemical solvents to dissolve the contaminants from the surface, we explore here a thermophysical method that could reinstate superhydrophilicity without changing the surface chemistry. Therefore, the novelty of this work lies in understanding how superhydrophilicity of substrates can be recovered by exploiting thermophysical mechanisms, like condensation. The experiments reveal that the combination of molecular desorption at the moving contact line and Laplace pressure differentials during condensate droplet nucleation and growth, causes the deposited contaminants to be removed -albeit not completely- from the surface. After exposing surfaces with different texture topologies to the ambient atmosphere for over eight months, we report convincing evidence that condensation can be more effective in reinstating superhydrophilicity than rinsing with acetone, ethanol, or water. Thus, a new path for scalable and environment-friendly wettability recovery opens up, by establishing surface microstructure design guidelines promoting durable superhydrophilic surfaces.

## 2. EXPERIMENTAL
### 2.1. Laser-etching process
*2.1.1. Sample preparation*
SAE 304 stainless steel (2 mm thickness) plates were purchased from McMaster-Carr. The plates comprised of a mirror-finished side, which was safely protected by a film. First, the plates were sheared into 3 cm × 3 cm pieces and the protective film was removed. Then, the samples were sonicated first in deionized water and subsequently in acetone (1 hour each), to remove any contaminants from the surface. Next, the samples were subjected to laser ablation.

For the scaled-up experiment, a plate of 18 cm × 26 cm dimensions was sized following the same procedure.

*2.1.2. Laser ablation*

The sample plates were surface-textured via ablation using a 1064 nm nanosecond pulsed 20-Watt Yb:fiber laser (TYKMA Electrox EMS 300) with spot diameter of 50 μm. The laser parameters used in this study are listed in **Table S1**. The line spacing (i.e. distance between two adjacent beam raster lines) was the main operational parameter, due to its strongest effect on the micro/nanomorphology of the created surface patterns.[24] Furthermore, laser-line spacing can affect the cost of the process, as smaller spacings extend the time required for the laser to ablate an area. Each substrate was treated twice: first in one direction and then in another, perpendicular to the original direction. This resulted in an orthogonal cross-hatched microstructure morphology on the surface (**Figure 1**). The sample plates in this work were labeled as SS75, SS125, and SS150, in accordance of the line spacing used to make them (75 μm, 125 μm, and 150 μm, respectively). For the scaled-up experiment, the larger sample (26 cm × 18 cm) was ablated with 150 μm laser-line spacing.

## 2.2. Sample cleansing process

The cleansing of the surfaces was attained by condensation performed inside an ESPEC SH-641 environmental chamber. The samples were attached to a cold plate (maintained at 10 °C) placed vertically inside the chamber to facilitate the downward flow of the condensed water due to gravity. The temperature of the chamber was set to 40°C and the relative humidity to 80%. Each testing cycle lasted for 1 hour, after which the surfaces were dried out using pure $N_2$ gas before being characterized. The condensation process was also captured on video using a Canon EOS T1i camera (Supporting Information video named "Condensation"). The amount of water collected during one condensation cycle and the collection rate were also measured, with the results presented in **Table S2**.

For the rinsing experiment, a Harvard Apparatus PHD ULTRA Syringe Pump was used to create a jet of acetone, ethanol, or water, with a flow rate of 50 mL/min (jet velocity 0.3 m/s). The plates were positioned at a 45° angle to maximize the contact with the jet and facilitate the downward flow of the liquid (Supporting Information video named "Rinsing"). After the rinse, the surfaces were dried out using pure $N_2$ gas and were subsequently characterized.

## 2.3. Water droplet spreading rate measurements

The spreading rate was quantified by releasing a 4.7 μL water droplet on the prepared surfaces through a syringe (BD) at very low Weber numbers $\sim O(10^{-2})$. The process was recorded at 1000 fps using a Phantom Miro M310 high-speed camera and Phantom Camera Control software. The spreading rate encompasses both the Washburn and Tanner regime (details in section 3 of Supporting Information), thus providing a comprehensive understanding of the transition from capillary-driven infiltration to dissipation-governed spreading, where viscous effects and contact line dynamics dictate the motion. The original wicking curves are presented in **Figure S1**. Particular care was taken to eliminate any lingering inertial or neighboring wind effects.

## 2.4. X-ray Photoelectron Spectroscopy (XPS)

The XPS measurements were performed on the prepared surfaces using a Thermo Fisher Nexsa G2 instrument. Analysis and peak fitting for C, O, and Fe elements were carried out using the Thermo Avantage software. All spectra (**Figures S2-S4**) were calibrated based on the C-C bond binding energy at 284.8 eV.

## 2.5. Scanning Electron Microscopy - Energy Dispersive Spectroscopy (SEM-EDS)

High-resolution images and elemental composition of the prepared surfaces were acquired by a JEOL JSM-IT500HR instrument. The results are presented in **Table S3**.

## 2.6. Nanotopology measurements

The topological characteristics of the surface nanofeatures were studied using the nanoTOPO-SEM software (Nanometrisis, Greece). Using the acquired high-magnification SEM images, the software analyzes the surface morphology and generates a collection of metrics characterizing the spatial aspects of the surface features (**Figures S5 and S6**).

## 3. RESULTS AND DISCUSSION

### 3.1. Pattern Morphology

An orthogonal cross-hatch pattern texture was applied on the stainless steel plates. Parallel trenches were ablated onto the surface with gaps corresponding to a prescribed line spacing. Thereafter, another set of parallel laser ablation lines perpendicular to the first set were applied on the surfaces. The fine-scale morphology of the created patterns can be seen in the SEM images of **Figures 1A, 1E, 1I**. As expected, the laser acts as a "plow" creating channels of ~50 μm width (consistent with the laser beam spot diameter), while the displaced material is primarily redeposited in between these microchannels. Depending on the line spacing used, the area in between the microchannels appears either as narrow peaks for the SS75 plate or wide plateaus for the SS125 and SS150 plates. These features can be more clearly observed in the optical profilometry images, where a 3D surface morphology is displayed (**Figures 1D, 1H, 1L**). As revealed in the magnified SEM images (**Figures 1B, 1F, 1J**), after laser ablation, the surfaces are covered with a layer of nanofeatures forming very distinctive repeatable topologies (**Figures 1C, 1G, 1K**).

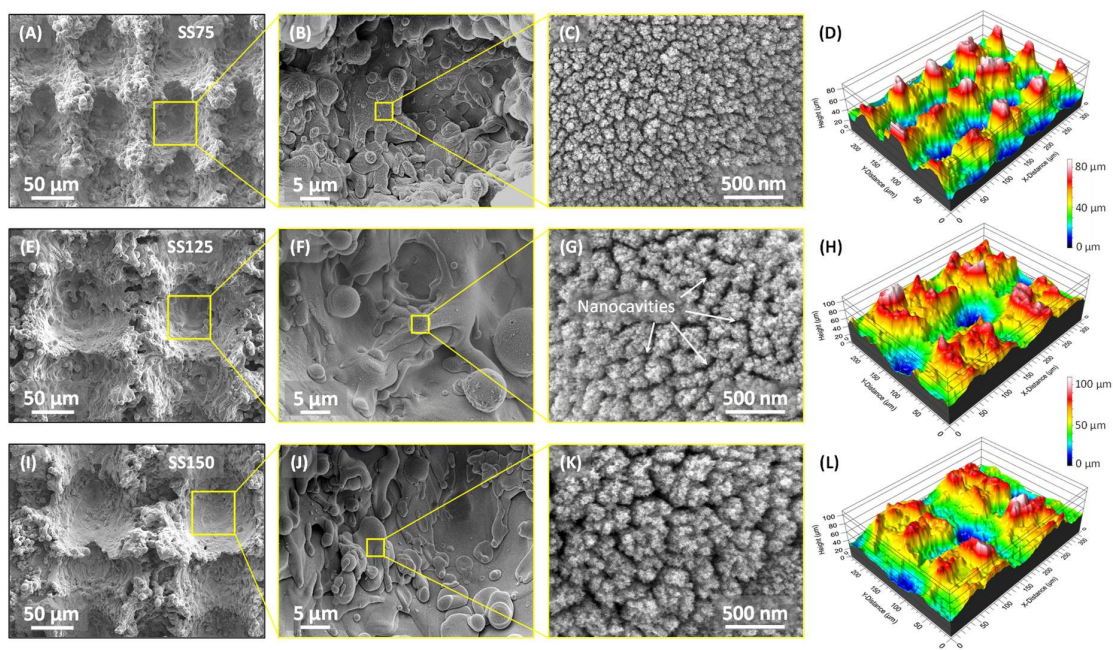

**Figure 1. Morphology of prepared surfaces.** SEM (left) and profilometry (right) images of SS75 (A-D), SS125 (E-H), and SS150 (I-L) plates after laser ablation. The profiles in (H) and (L) share the same color scale bar (0 - 100 μm), while (D) has a separate scale (0 - 80 μm).

This layer has been previously identified to consist of metal oxide nanoparticles,[42,43] produced by oxidation induced by the laser beam and consequent re-deposition during laser ablation.[44-46] Indeed, the EDS results (**Figure 2**) reveal the significant increase of oxygen from 2.8% (at.) on an unprocessed stainless steel plate to 25-27% on the laser-ablated plates. Interestingly, the nanofeature topology seems to be related to the line spacing. Specifically, the SS75 plate shows the highest abundance of densely packed nanofeatures, while the SS125 and SS150 plates display a sparser distribution.

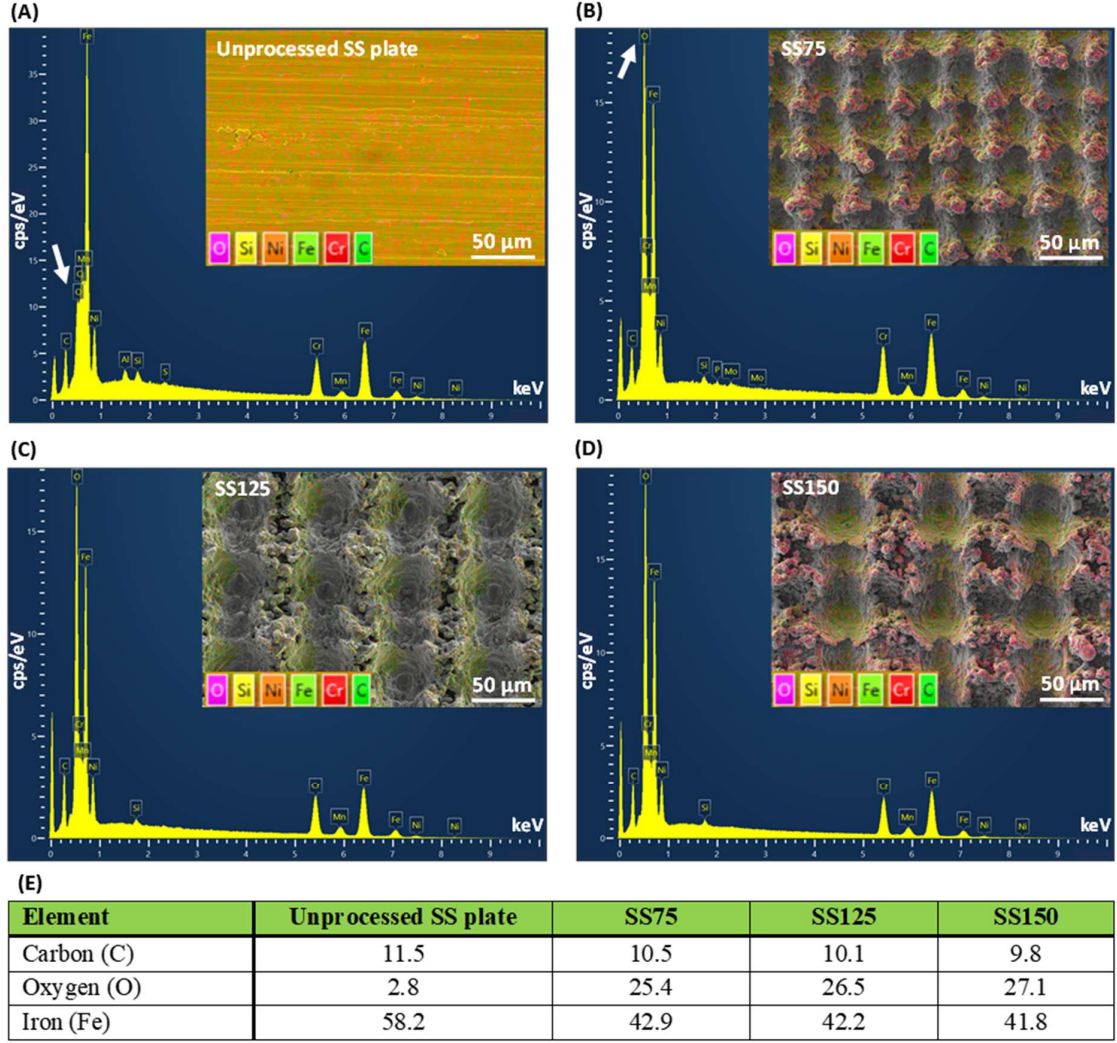

**Figure 2. Surface chemistry of prepared surfaces.** (A-D) EDS spectra and mapping of the unprocessed stainless steel plates (A), SS75 (B), SS125 (C), and SS150 (D). (E) Atomic percentage of C, O, and Fe for the unprocessed and laser-ablated plates, as measured by EDS.

### 3.2. Characterizing Superhydrophilicity after Laser Ablation

The laser-ablated surfaces were superhydrophilic, as total wetting was observed when a 4.7 μL droplet of deionized water was placed at the center of the substrate, with minimal impact velocity, Weber number $\sim O(10^{-2})$, to avoid inertial spreading effects. It is well established that the spreading parameter ($S$) dictates whether a surface gets wetted by a liquid droplet[1]

$$S = \gamma_{SV} - (\gamma_{SL} + \gamma_{LV}) \qquad (1)$$

where $\gamma_{SV}$, $\gamma_{SL}$, and $\gamma_{LV}$ denote the surface energies of the solid-vapor, solid-liquid, and liquid-vapor interfaces, respectively. $S \geq 0$ predicts complete wetting of the substrate by the liquid. Earlier research of droplet spreading behaviors on high-energy substrates has shown that the relevant velocity of the contact line ($U$) can be modeled from the apparent contact angle ($\theta_a$) of the liquid on the surface, under the assumptions of low Bond number (ratio of the gravitational forces over the surface tension forces), low capillary number,[47] and the droplet having the shape of a spherical cap.[48,49] These assumptions render the problem one-dimensional, where variations in contact-line velocities at different points along the wetting front can be accounted for by quantifying the wetted area ($A$) in place of the wetted radius ($R$). Therefore,

$$U = \frac{dR}{dt} \approx \frac{\gamma_{LV}}{\eta} \theta_a^m \quad (2)$$

$$\theta_a \approx \left[ \frac{\eta}{2\gamma_{LV}} \sqrt{\frac{1}{\pi A}} \frac{dA}{dt} \right]^{1/m} \quad (3)$$

where $\eta$ is the liquid viscosity, and $m = 3$ for liquid spreading (S > 0) on a dry surface.[48] Moreover, analysis of the developed microstructures must be carried out in order to distinguish whether the spreading arises from capillarity or hemiwicking through the microstructures.[50] Assuming that the Young-Laplace equation $\gamma_{LV} \cos\theta = \gamma_{SV} - \gamma_{SL}$ is valid at the contact line, and the top of the microstructures remain dry during liquid imbibition, the critical angle for imbibition ($\theta_c$), can be calculated from[51]

$$\cos\theta_c = \frac{1-\phi_S}{r_f-\phi_S} \quad (4)$$

where $r_f$ denotes the Wenzel roughness,[1] defined as the ratio of the actual surface area to the projected surface area, and $\phi_s$ is the projected solid area fraction that remains dry during the liquid spreading in between the surface asperities.[51] The corresponding values of the above quantities along with the average surface roughness ($S_a$) and intrinsic contact angle ($\theta_i$) for the three surfaces are listed in **Table 1**. For values of $\theta_a > \theta_c$, the Wenzel state is attained, where the droplet spreading arises completely due to capillarity through the microstructures. Thus, the intrinsic contact angle ($\theta_i$) for the same material with $r_f = 1$ can be estimated from $\cos\theta_a = r_f \cos\theta_i$. However, in cases where $\theta_a < \theta_c$, hemiwicking ensues and an alternate expression, namely $\cos\theta_a = 1 - \phi_s(1 - \cos\theta_i)$ must be employed. Both these expressions arise from the hypothesis that for extremely smooth surfaces (where all effects of physical interaction of the liquid with the surface can be neglected) the dynamic contact angles can be assumed equal to the respective static contact angles. Furthermore, estimation of the intrinsic angles on such smooth surfaces (for $r_f=1$) allows for monitoring how the material surface chemistry changes with VOC adsorption, while neglecting the additional contact-line dynamics or hemiwicking effects arising due to surface micro/nano-textures.

**Table 1.** Surface metrics and contact angles for the three SS textures employed in this study.

| Sample | Measured parameters | | | | | |
|---|---|---|---|---|---|---|
| | $\phi_s$ | $r_f$ | $S_a$ (μm) | $\theta_a$ (º) | $\theta_c$ (º) | $\theta_i$ (º) |
| SS75  | 0.3 ± 0.03 | 1.9 ± 0.1 | 12.9 ± 0.8 | 3.1 ± 0.3 | 62.2 ± 0.1 | 55.7 ± 0.8 |
| SS125 | 0.2 ± 0.01 | 1.7 ± 0.1 | 14.3 ± 0.4 | 3.1 ± 0.3 | 57.5 ± 0.1 | 55.7 ± 0.9 |
| SS150 | 0.1 ± 0.01 | 1.6 ± 0.1 | 15.8 ± 0.4 | 3.2 ± 0.2 | 52.8 ± 0.1 | 55.7 ± 0.5 |

$\phi_s$, $r_f$ and $S_a$ are determined from profilometry images

Therefore, extrapolation of the above theory to monitor the contact angle variation with time as VOCs adsorb on the textured surface allows estimation of the changes in the underlying physicochemical properties of the surface.[1,47] Since the prepared surfaces experience no change in $r_f$ during ambient exposure, $\theta_i$ reveals changes in their chemical properties (and thus, intrinsic wettability) due to VOC adsorption.

### 3.3. Superhydrophilicity Degradation during Ambient Exposure

The superhydrophilicity of the prepared 3 × 3 cm$^2$ surfaces was examined before and after exposure to a research laboratory environment in Chicago for 1 month (short-term) and 8 months (long-term). The VOCs in the ambient environment get deposited on the surface leading to superhydrophilicity degradation of the surfaces, which consequently results in decreased spreading rate (or hemiwicking).[28-30]

#### 3.3.1. Short-term Exposure

**Figure 3A** shows the spreading rate (mm$^2$/s) of a 4.7 μL water droplet deposited on each of the textured surfaces. Soon after fabrication, all the surfaces exhibit a spreading rate of ~400 mm$^2$/s, which can be attributed to the similar characteristic width (50 μm) of the created channels and thus, the exertion of similar capillary forces.[52] Our spreading rate aligns with the Lucas-Washburn relation, in agreement with other laser-textured metallic surfaces.[2,23,30,31]

After 1 week of open-air exposure in the laboratory, the water spreading rate decreases to 170 mm$^2$/s for the SS75 plate, 145 mm$^2$/s for SS125, and 125 mm$^2$/s for SS150. After week 3, a plateau is seemingly reached at 90 mm$^2$/s, 75 mm$^2$/s, and 65 mm$^2$/s, respectively. The superhydrophilicity degradation of the SS75 sample is presented in **Figure 3B** (and Supporting Information video named "Spreading") through the timelapse of a water droplet spreading on the surface before and after 1 month of ambient exposure. The observed degradation can be attributed to the gradual adsorption and accumulation of contaminants on the surface, a phenomenon well established in the literature.[29,41,53]

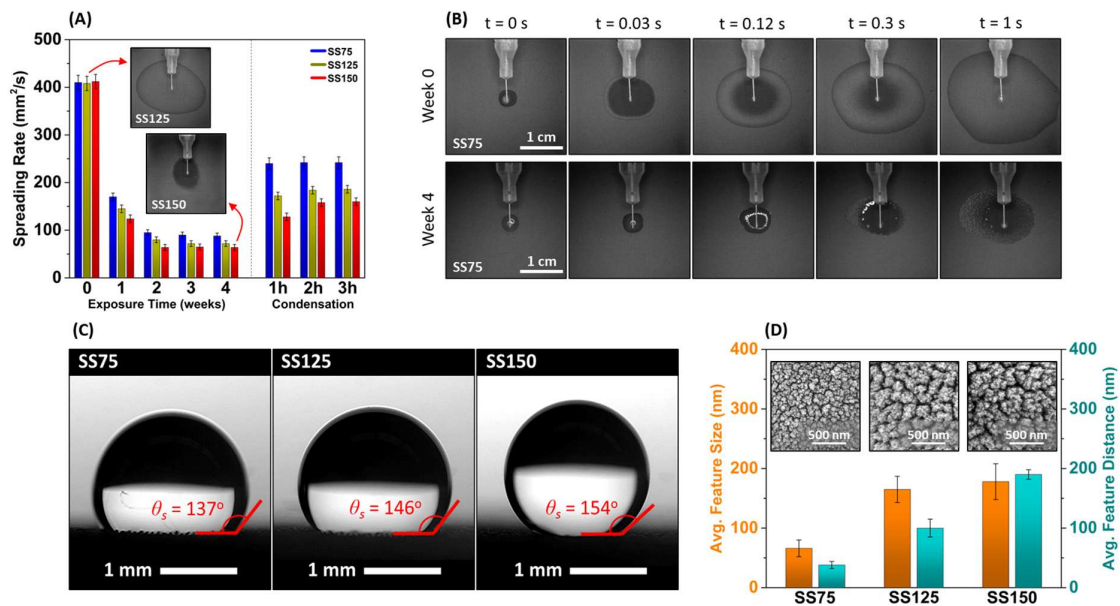

**Figure 3. Water spreading rate measurements.** (A) Spreading rate on SS75, SS125, and SS150 plates after short-term exposure in ambient environment and consecutive condensation cycles. (B) Timelapse of water droplet spreading on SS75 sample before and after 1 month of exposure in ambient air. (C) Static contact angles on the plates after long-term (8 months) exposure in ambient environment. (D) Average size of the formed nanofeatures and distance between them for SS75, SS125, and SS150 plates. The inset SEM images show respective characteristic areas of nanofeatures.

XPS measurements were performed on the prepared surfaces before and after exposure to the laboratory environment. The C1s spectra (**Figure S3**) show three peaks at 284.8 eV, 286.4 eV, and 288.6 eV, attributed to C-C(H) from hydrocarbon chains or graphitic structure, the alcohol/ether C-O groups, and the aldehydes/ketones C=O groups, respectively.[54,55] The O1s spectra (**Figure S4**) show peaks at 529.8 eV, 531.4 eV, and 533.1 eV, attributed to the Fe-O bonds, the alcohol/ether C-O groups, and the aldehydes/ketones C=O groups, respectively.[54,55] **Table 2** shows the carbon (C1s) and oxygen (O1s) content relative to iron (Fe2p) at different stages of the experiments. The ratio of C1s to Fe2p, which indicates the amount of adsorbed organics,[56] increased after 1 month of exposure by 80% for the SS75 plate, 92% for SS125, and 100% for SS150. Similarly, the O1s to Fe2p ratio increased by 20%, 20%, and 22%, respectively. The XPS results of the degraded surfaces are consistent with VOCs in the Chicago area atmosphere,[57,58] and directly correlate with the observed decrease of spreading rate. As expected, this change of chemical properties due to VOCs adsorption is also reflected on the $\theta_i$ values, which increase after exposure to the ambient environment and then decrease (as superhydrophilicity is recovered) when undergoing condensation from humid air (see later sections). The variation in extent of $\theta_i$ increase indicates that VOCs adsorption is more prominent on the SS150 plate as opposed to SS75 (**Table 2**). This trend, observed also during the long-term ambient exposure, is attributed to the morphology (or dense packing) of the nanoparticles on the surface.[41]

**Table 2.** C/Fe and O/Fe ratios measured by XPS and intrinsic contact angles ($\theta_i$) for the laser-ablated plates at different stages of the ambient exposure experiments.

| Stage | SS75 | | | SS125 | | | SS150 | | |
|---|---|---|---|---|---|---|---|---|---|
| | C/Fe | O/Fe | $\theta_i$ (º) | C/Fe | O/Fe | $\theta_i$ (º) | C/Fe | O/Fe | $\theta_i$ (º) |
| Week 0 | 1.15 | 2.98 | 55.7 ± 0.8 | 1.22 | 2.92 | 55.7 ± 0.9 | 1.18 | 2.84 | 55.7 ± 0.5 |
| Week 4 | 2.08 | 3.56 | 58.5 ± 0.4 | 2.34 | 3.42 | 59.8 ± 0.5 | 2.36 | 3.44 | 60.9 ± 0.5 |
| After condensation | 1.60 | 2.66 | 56.7 ± 0.4 | 1.91 | 2.95 | 58.5 ± 0.5 | 1.98 | 3.03 | 59.2 ± 0.4 |
| Recovery* (%) | - | - | 64.3 | - | - | 31.7 | - | - | 32.7 |

*Recovery of intrinsic surface wettability (Eq. 5)

Thereafter, the recovery of intrinsic wettability (towards reinstating the smooth-surface chemical properties) was estimated by

$$Recovery = 1 - \frac{\theta_i^{After\ condensation} - \theta_i^{Week\ 0}}{\theta_i^{Week\ 4} - \theta_i^{Week\ 0}} \quad (5)$$

thus, offering a quantitative criterion for this recovery.

### 3.3.2. Long-term Exposure

After 8 months of ambient exposure, the surfaces completely lose their superhydrophilicity and hemiwicking properties, thereby exhibiting typical hydrophobic behavior. The deposited water droplet no longer spreads on the surface, and instead forms a high static contact angle ($\theta_S >$ 120°) for all plates (**Figure 3C**). Specifically, plate SS75 displays a $\theta_S = 137° ± 9°$, SS125 $\theta_S = 146° ± 3°$, and SS150 $\theta_S = 154° ± 2°$ (crossing the widely accepted threshold of superhydrophobicity). It is essential to note that the degree of hydrophobicity is not dependent on the $r_f$ value of the surfaces -which increases with the decreasing line spacing of the laser ablation- but on the level of VOC adsorption from the environment. Furthermore, the amount of VOCs adsorbed depends on the nanoparticle morphology (size and density) on the top layer of the textured surfaces.

### 3.4. Examination of Surface Topology

The presence and characteristics of surface nanofeatures play a crucial role in the superhydrophilicity degradation (or its reinstatement), as they can dictate the adsorption of contaminants and their diffusion on the underlying surfaces.[41,59] Therefore, the formed nanotopology was examined, and the results are presented in **Figure 3D**. As seen in the graph, the average size of nanofeatures and distance between them greatly varies depending on the beam raster-line spacing. Specifically, the average nanofeature size was calculated (complementary image analysis from SEM micrographs presented in **Figures S5** and **S6**) to be 66 nm, 165 nm, and 178 nm, for SS75, SS125, and SS150, respectively.[60] Similarly, the average distance between these nanofeatures was 38 nm, 100 nm, and 190 nm, respectively. The created nanocavities appear to be characteristic and directly correlated to the superhydrophilicity degradation results.[41] We hypothesize that the smaller gaps observed on the SS75 plate significantly inhibit the diffusion of contaminants to the surface, thus slowing down the degradation caused by the accumulated organics. On the other hand, as the gap between the nanofeatures increases for the SS125 and SS150 samples, higher accumulation of contaminants occurs, resulting in faster wettability degradation. This argument is in-line with our hypothesis that a greater nanoparticle density delays adsorption of VOCs from the ambient environment.

### 3.5. Attempting to Reinstate Superhydrophilicity via Rinsing

Earlier exploits into surface cleaning have often adapted rinsing (or washing) the surface with solvents to dissolve the adsorbed VOCs. While some approaches even employ ultrasonication of the substrates in the solvents for increased efficacy, we argue that such an approach would not only be impractical but also infeasible for bigger substrates. Furthermore, ultrasonication is well-known for damaging or even removing nanoparticles altogether from the surface, and therefore it is not suited in our scenario, where the major objective is to revive the surface wickability while retaining its micro/nanomorphology. Therefore, two methods were adopted for testing the efficiency of VOC removal from the surface: rinsing and condensation. Rinsing is highly practical owing to its simplicity, and therefore we washed the surface with common solvents like acetone, ethanol, and deionized water (Supporting Information video named "Rinsing").

For long-term exposed plates (**Figure 4B**) rinsing with acetone and ethanol barely improved the wettability for all three samples. Rinsing the surface with deionized water was found to be more effective, as the water contact angle $\theta_S$ on the SS75, SS125 and SS150 plates decreased to 85°, 100°, and 114°, respectively, after 2 hours of rinsing. More extended rinsing (total of 8 hours) of the substrates with water followed, where all surfaces turned hydrophilic (15° < $\theta_S$ < 90°). Interestingly, even such very long durations of washing failed to reinstate the superhydrophilicity and hemiwicking behavior of the surfaces.

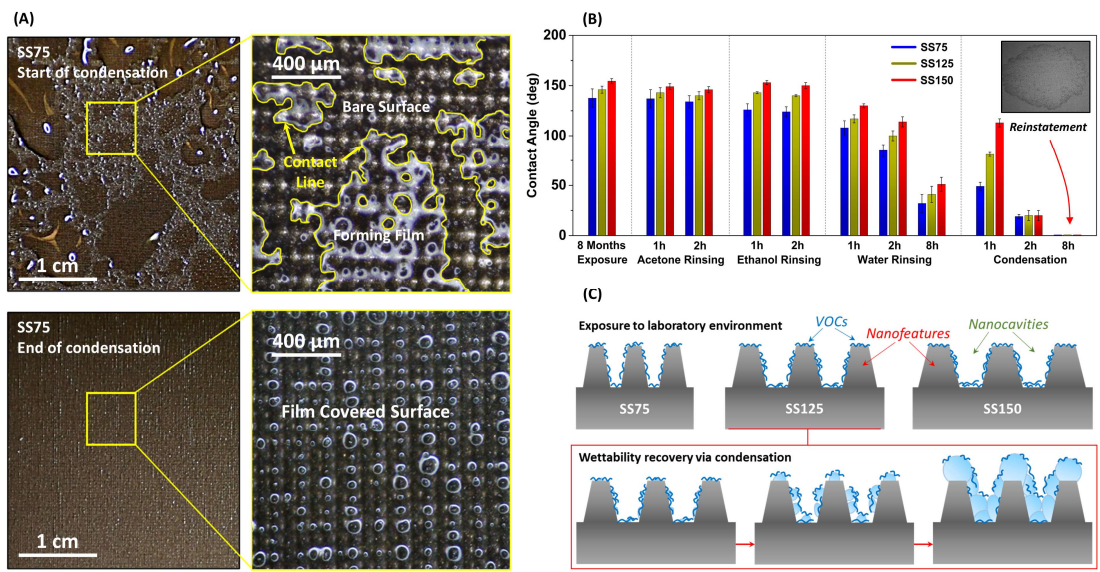

**Figure 4. Wettability recovery via condensation.** (A) High-magnification imaging of contaminated SS75 plate during condensation showing the contact line (in yellow) movement. (B) Static contact angles of water droplet on SS75, SS125, and SS150 plates after rinsing with acetone, ethanol, water, and after condensation. After 8h of condensation, there is complete reinstatement of superhydrophilicity. (C) Simplified schematic (top) of the surface nanofeatures and contaminant accumulation after exposure to ambient environment. Proposed mechanism (bottom) of wettability recovery via condensation (example drawn for the SS125 plate only).

## 3.6. Reinstating Superhydrophilicity via Condensation

Condensation from warm humid air in an environment-controlled chamber (where a prescribed ambient temperature and relative humidity can be set and maintained) was employed in exploration of a non-chemical, environment friendly, and scalable methodology for effective removal of contaminants from the ambient-exposed surfaces. The chamber temperature and relative humidity were set to 40 °C and 80% respectively, whereas the substrate was maintained at 10 °C with the help of an attached cold plate (Section 2 of Supporting Information).

Due to the presence of superhydrophilicity inhibiting VOCs on the surfaces after ambient exposure, condensation occurred both in dropwise and filmwise modes (**Figure 4A**). After an hour-long condensation cycle (measured from the moment when temperature and humidity reached the set values), the wetting ability of the substrates that were short-term exposed was partially recovered. As seen in **Figure 3A**, spreading rates of 240 mm$^2$/s, 185 mm$^2$/s and 150

mm$^2$/s were measured for the SS75, SS125 and SS150, respectively. It is expected that the superhydrophilicity will keep on improving as the contact-line movement causes further recovery of the original surface properties via VOC removal.

In the case of long-term ambient exposed plates (**Figure 4B**), an hour-long condensation cycle decreased the $\theta_S$ of SS75, SS125 and SS150 plates to 49°, 81°, and 113°, respectively. A subsequent condensation cycle (1 hour) further improved the wettability, as the $\theta_S$ of the plates reached 19°, 20°, and 20°, respectively. Therefore, this approach was adopted and condensation on the samples was continued for 6 more hours (total condensation time of 8 hours) until their superhydrophilicity and hemiwicking nature was successfully reinstated. Specifically, the final spreading rate was measured to be 275 mm$^2$/s for the SS75 plate, 184 mm$^2$/s for SS125, and 105 mm$^2$/s for SS150. Albeit these rates were not comparable to the ones observed before ambient exposure, they were comparable to those observed for the short-term exposed surfaces. Since the cleaning efficiency is dependent on the adhesion between the contaminants and the surface, it is possible that the lack of full recovery can be attributed to chemisorbed contaminants, as both physisorption and chemisorption can occur after lengthy ambient exposure. While "physisorption" refers to the physical adsorption of pollutants onto a surface due to weak forces -like Van der Waals interactions- "chemisorption" involves the formation of chemical bonds between the organic contaminants and the functional groups of the surface.[61,62] Thus, even organic solvents, such as acetone or ethanol, that can wash off physisorbed contaminants are unable to remove those chemically adsorbed onto a surface through molecular bonds.[63,64] An in-depth study of superhydrophilicity reinstatement after physisorbed vs. chemisorbed VOCs is beyond the scope of this work.

### 3.7. Mechanism for Wettability Recovery

We hypothesize that the recovery in surface superhydrophilicity is attributed to two factors: 1) The nanofeatures serve as preferential nucleation sites during condensation, allowing substantially more time for the contaminants to dissolve into the formed water droplets before they flow downwards due to gravity; 2) The moving contact line during condensation can desorb physically adsorbed molecules from inside the nanocavities, enabling "in-depth" cleaning. On the other hand, when simple rinsing is used, the liquid is in contact with the contaminants for a limited amount of time and therefore only the more exposed and easier to dissolve contaminants (i.e., short-chain organics) get removed.

Earlier experiments of molecular desorption dynamics at the contact line have tied the desorption phenomena predominantly to movement of the liquid interface over the solid surface.[65] High-magnification imaging of the contaminated surfaces during condensation from warm humid air shows movement of the contact line (**Figure 4A** and Supporting Information video named "Condensation") before a uniform film is formed on the substrate surface. For extremely low capillary numbers (when the contact line either advances or recedes on the surface), such absorbed molecules can be modelled as local chemical heterogeneities. An estimation of the capillary energy for a moving (dynamic) contact line on such local physisorbed surface heterogeneities showed it to be of the order of ~7 $k_BT$ ($E_{cap}$), which is comparable to the absorption energy of ~10 $k_BT$ per molecule ($k_B$ Boltzmann constant, $T$ temperature).[65] This capillary energy further increases for smaller perimeters of the contact line, thereby supporting the hypothesis that the nanoscale features are essential for the "self-cleaning" nature of the laser-textured surfaces. Therefore, it is feasible that the contact-line motion provides sufficient energy for the adsorbed molecules to detach from the surface and enter the condensate bulk. Furthermore, we propose that the effect of desorption is magnified when a condensate molecule grows inside a nanostructure pore and gets released on the surface. There exist substantial Laplace pressure gradients (noting that the Laplace pressure is inversely proportional to the radius of curvature) from inside the nanopore to millimeter scale condensate droplets on the surface, which can create significant diffusion gradients for the VOCs to flow out from the micro/nanopores into bigger condensate drops on the surface (**Figure 4C**). The abundance of nanofeatures leads to more nucleation sites during condensation, which allows the generation of more nanodroplets that can participate in the removal of contaminants. This is also evident by the collected amount of condensed water in **Table S2**, which is higher for the SS75 plate, and lower for SS125 and SS150. The nucleated nanodroplets experience substantial Laplace pressure differences, which create a strong gradient for adsorbed contaminants to flow out from the surface into the condensate, thus facilitating the observed superhydrophilicity recovery. This is highlighted by the XPS results in **Table 2**, where the ratio of C1s to Fe2p after condensation is reduced by 24% for the SS75 plate, 18% for SS125, and 16% for SS150 (detailed XPS spectra in **Figures S2-S4** of Supporting Information). Importantly, these values correlate with the observed recovery of spreading rate, with SS75 showing the best recovery among the prepared surfaces.

### 3.8. Ramifications for Scaled-up Applications

The scalability and robustness of the proposed methodology was further tested on a plate with much larger area (26 cm × 18 cm) than the original square samples. To ensure that the contamination results are not specific to the Chicago area, the plate was exposed to a research laboratory environment in Houghton, Michigan (Michigan Technological University) for 4 months. Afterward exposure, the sample was shipped back to Chicago for further testing. Goniometry measurements showed that after ambient exposure, the plates exhibited an average $\theta_s$ of 126° ± 9°, with superhydrophilicity completely absent from any part of the surface. Thereafter, the sample was placed on top of a cold plate maintained at 2°C, at typical room conditions (temperature 24-25°C, relative humidity 38-40%) and condensation was observed (**Figure 5A**). During the process, the ability to maintain a uniform condensation film was tested by drying locally the surface with a jet of pure $N_2$ (99.999%). As seen in the Supporting Information video named "Condensation-RT", the condensate film quickly recovers to completely cover the substrate, indicating the robustness of the method. After 30 hours of continuous condensation, the sample was detached from the cold plate, dried with $N_2$, and the hemiwicking properties were characterized with high-speed imaging. Similar to the smaller samples, the superhydrophilicity was fully reinstated after condensation with the plate showing a spreading rate of 295 mm$^2$/s (**Figure 5B**).

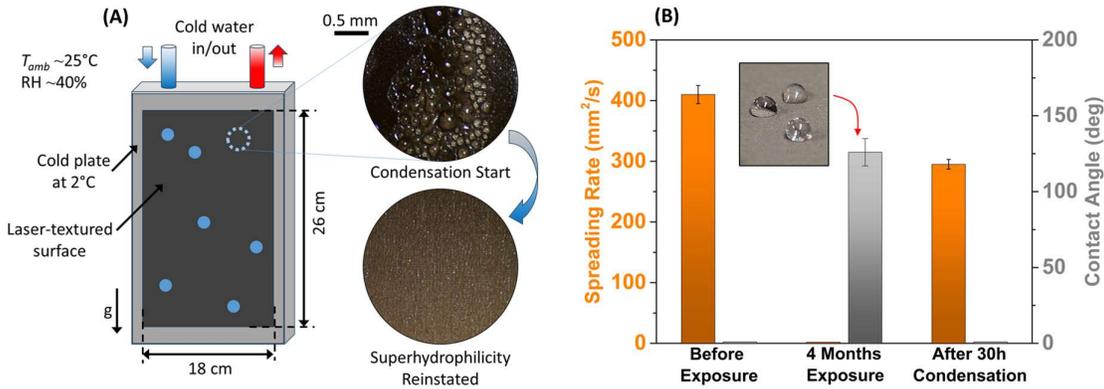

**Figure 5. Scaled-up application.** (A) Schematic of the sample mounting setup used for evaluating the scalability of the condensation method for reinstating superhydrophilicity. The circular images at center show the dropwise condensation at the beginning of the process, and the reinstatement of superhydrophilicity (formation of liquid film) after a few hours of continuous condensation. (B) Spreading rate and contact angle values measured before exposure in ambient environment, after ambient exposure for 4 months, and after 30h of continuous condensation.

These results prove the scalability of the condensation cleansing method and highlight that: 1) The approach is effective, regardless of the geographical location where a surface is exposed; 2) There is no need for extreme condensation conditions, as normal (room) temperature and humidity values would suffice.

## 4. CONCLUSIONS

Superhydrophilic stainless steel surfaces were produced by surface texturing using nanosecond pulsed laser ablation. The superhydrophilicity degradation due to accumulation of VOCs from ambient exposure and the subsequent reinstatement via condensation from humid air were studied. It was revealed that the nanofeature topology has a significant impact on the wettability loss of the surfaces and its recovery. The results indicate that the formation of narrow nanocavities inhibits the diffusion of contaminants and prevents their accumulation on the underlying solid. When these nanocavities are wider, the deposited organic layer penetrates onto the underlying surface, thus leading to accelerated degradation and less efficient recovery. Besides controlling the deposition of the VOCs on the underlying surface, the nanofeatures also significantly affect the recovery through condensation. The nucleating nanodroplets on the nanofeatures experience substantial Laplace pressure differences, which ultimately create a strong gradient for contaminants to be removed from the surface into the condensate. Furthermore, the moving contact line desorbs molecules that were physically adsorbed to the surface, facilitating VOC removal and therefore "cleansing" at a nanoscale level. Due to these phenomena, condensation proves to be more efficient than chemical rinsing with acetone, ethanol, or water, and thus emerges as a promising scalable and environment-friendly alternative methodology for reinstating the hemiwicking on a textured superhydrophilic surface, without altering its micro/nanosurface morphology.

## ASSOCIATED CONTENT

**Supporting Information – (see https://doi.org/10.1021/acsami.5c00724 )**

Detailed experimental methods, XPS measurements, SEM-EDS analysis, large scale test (PDF)

Video of the superhydrophilicity degradation ("Spreading")

Video of the rinsing process ("Rinsing")

Video of the contact line movement during condensation ("Condensation")

Video of the condensate film recovery at typical room conditions ("Condensation-RT")

## AUTHOR INFORMATION

**Author Contributions**

I.P. performed the spreading rate measurements and the condensation experiments. A.M. performed the SEM-EDS analysis, the profilometry analysis, and the laser-ablation of the plates. A.P. performed the nano-topology measurements and helped with the laser-ablation process and spreading rate measurements. S.N. performed the XPS analysis. C.M.M.

supervised the work and verified the experimental results. I.P., A.M., A.P., and C.M.M. contributed to writing the manuscript.

**Notes**

The authors declare no conflict of interest.

**ACKNOWLEDGEMENTS**

This study was funded by the U.S. National Science Foundation (Award 2133229 to Michigan Technical University, and a subcontract to UIC). The authors acknowledge the use of the facilities at the University of Illinois Chicago - Research Resources Center, namely the Nanotechnology Core Facility, Electron Microscopy Core and the Scientific Instrument Shop for help with sample preparation, characterization, and design of the experiments, respectively. The authors thank Prof. Raymond Shaw for discussions, and acknowledge the assistance from Grant Schlaff and other collaborators at the Pi cloud chamber (Michigan Technological Univ.) with the ambient exposure of the larger sample. The topological characteristics of the surface nanofeatures were determined using the nanoTOPO-SEM software (Nanometrisis, Greece).

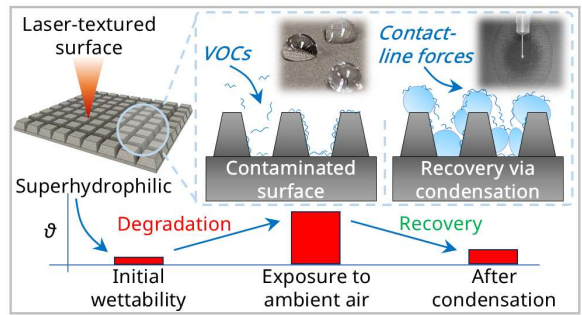